\documentclass[prl,twocolumn,showpacs,preprintnumbers,amsmath,amssymb,superscriptaddress]{revtex4}
\usepackage{dcolumn}% Align table columns on decimal point
\usepackage{bm,graphicx}% bold math

\begin{document}

\title{Carrier scattering from dynamical magneto-conductivity \\ in quasi-neutral epitaxial graphene}
\author{M. Orlita}\email{milan.orlita@lncmi.cnrs.fr}
\affiliation{Laboratoire
National des Champs Magn\'etiques Intenses, CNRS-UJF-UPS-INSA, 25, avenue des
Martyrs, 38042 Grenoble, France}
\affiliation{Institute of Physics, Charles University in Prague, Czech Republic}
\author{C. Faugeras}
\affiliation{Laboratoire National des Champs Magn\'etiques Intenses,
CNRS-UJF-UPS-INSA, 25, avenue des Martyrs, 38042 Grenoble, France}
\author{R. Grill}
\affiliation{Institute of Physics, Charles University in Prague, Czech Republic}
\author{A. Wysmolek}
\affiliation{Institute of Experimental Physics, University of
Warsaw, Ho\.{z}a 69, PL 00-681 Warsaw, Poland}
\author{W. Strupinski}
\affiliation{Institute of Electronic Materials Technology, PL 01-919 Warsaw,
Poland}
\author{C. Berger}
\affiliation{School of Physics, Georgia Institute of Technology, Atlanta,
Georgia 30332, USA} \affiliation{Institut N\'{e}el/CNRS-UJF BP 166, F-38042
Grenoble Cedex 9, France}
\author{W. A. de Heer}
\affiliation{School of Physics, Georgia Institute of Technology, Atlanta,
Georgia 30332, USA}
\author{G. Martinez}
\affiliation{Laboratoire National des Champs Magn\'etiques Intenses,
CNRS-UJF-UPS-INSA, 25, avenue des Martyrs, 38042 Grenoble, France}
\author{M. Potemski}
\affiliation{Laboratoire National des Champs Magn\'etiques Intenses,
CNRS-UJF-UPS-INSA, 25, avenue des Martyrs, 38042 Grenoble, France}
\date{\today}

\begin{abstract}
The energy-dependence of the electronic scattering time is probed
by Landau level spectroscopy in quasi neutral multilayer epitaxial
graphene. From the broadening of overlapping Landau levels we find
that the scattering rate $1/\tau$ increases linearly with energy
$\epsilon$. This implies a surprising property of the Landau level
spectrum in graphene -- the number of resolved Landau levels
remains constant with the applied magnetic field. Insights are
given about possible scattering mechanisms and carrier mobilities
in the graphene system investigated.

\end{abstract}


\maketitle

Understanding carrier scattering in graphene~\cite{GeimScience09} is one of the
fundamental but still unclear issues in the physics of electronic
properties of this noble
material. Initial, model prediction for graphene~\cite{ShonJPSJ98,AndoJPSJ02} done for
short- and long-range scatterers independent of carrier density assumed the relaxation
rate $1/\tau$ that is directly proportional to carrier energy $\epsilon$.
This is a simple consequence of the linear density of states with
energy, $D(\epsilon)\sim|\epsilon|$, in graphene which displays the
Dirac band structure with a constant Fermi velocity $v_{F}$. The
$1/\tau$$\sim$$|\epsilon|$ relation implies that the low temperature
(Boltzmann) conductivity $\sigma \sim v_{F}^{2} \tau(E_{F})D(E_{F})$ of
graphene is insensitive to changes of carrier concentration $n$
and/or position of the Fermi energy $E_{F} \sim \sqrt{|n|}$
\cite{ShonJPSJ98,AndoJPSJ02}. Surprisingly, experiments on gated
graphene flakes show different behavior, the conductivity varies
linearly with carrier density: $\sigma(n) \sim |n|$ \cite{NovoselovNature05,ZhangNature05}.
The simple $\tau(\epsilon) \sim |\epsilon|^{-1}$ rule may
not apply if dominant scatterers are of the specific resonant
type or ripples type~\cite{NiNanoLetters10,KatsnelsonPTRSA08},
as well as for charge impurities%~\cite{AndoJPSJ06}.
~\cite{AndoJPSJ06,NomuraPRL06,HwangPRL07,ChenNaturePhys08}.
Nevertheless,
the unexpected, strong dependence of $\sigma(n)$ on carrier
concentration remains puzzling~\cite{NiNanoLetters10,PonomarenkoPRL09} and
it is not fully clarified whether this behavior is intrinsic to graphene or has a
well-defined extrinsic origin.

In this Letter we present a positive test of the $1/\tau \sim
|\epsilon|$ rule, which we have performed on quasi-neutral layers of
epitaxial %~\cite{BergerJPCB04,SadowskiPRL06}.
graphene~\cite{BergerJPCB04,SadowskiPRL06,deHeerSSC07,deHeerJPD10}.
To probe this graphene system with a fixed Fermi energy we profit of
the dynamical magneto-conductivity method and follow the energy dependence
of the spectral broadening $\Gamma$ of inter Landau level (LL) transitions
measured in our infrared transmission experiments. Notably, the
optical conductivity response of graphene involves interband
excitations across the Dirac point and allows us to to determine
the parameter $\Gamma$ over a broad energy range. We have found that
in the limit of strongly overlapping LLs, which corresponds to the
quasi-classical limit in graphene and in which the simple relation $\Gamma=\hbar/\tau(\epsilon)$
is justified, this spectral broadening becomes $\Gamma=\alpha \cdot |\epsilon|$. This
linear dependence on energy is valid at all magnetic fields and
for all transition indexes ($\alpha=0.026$ in the sample
investigated). This experimentally determined scattering rate $1/\tau(\epsilon)\sim |\epsilon|$
dependence rules out specific extrinsic scattering mechanisms, such as resonant
scatterers or charge impurities, in the epitaxial graphene layers which are
protected from the environment. A straightforward but intriguing consequence of our
finding is that the number of resolved LLs is constant,
independent of the magnetic field.
Possible consequences of the
$1/\tau\sim|\epsilon|$ relation on the carrier mobility in
extrinsically unperturbed graphene are discussed.

The studied sample is a standard multilayer epitaxial graphene (MEG) grown on the C-terminated surface of silicon carbide
(4H-SiC[000\={1}])~\cite{BergerJPCB04} with intentional thickness of 50 layers. A significant
part of layers displays an electronic band structure as that of an isolated graphene monolayer, which
results from the characteristic rotational stacking of these graphene sheets~\cite{HassPRL08}. This is confirmed
by micro-Raman spectra measured on our sample, which show a single-component 2D
band~\cite{FerrariPRL06,FaugerasAPL08}, with some Bernal-stacked residuals on selected locations.
The sheets studied in this experiment are quasi-neutral~\cite{OrlitaPRL08II}, only several layers
close to the interface and on the surface of MEG become significantly doped (up to $10^{13}$~cm$^{-2}$)~\cite{LinAPL10}.

To measure the infrared transmittance, the sample was exposed to
the radiation of a globar, which was analyzed by a Fourier
transform spectrometer and delivered to the sample via light-pipe
optics. The transmitted light was detected by a composite
bolometer which operated at $T=2$~K and which was placed directly below
the sample. Measurements were done in a Faraday
configuration, using a superconducting magnet. The
sample is fully opaque in the energy range 85-120~meV and also
displays a relatively weak transmission around 200~meV. Both
absorption bands are related to phonon-related absorption in the
SiC substrate. The spectra presented here are either conventional
magneto-transmission spectra normalized by the zero field
transmission $T(\omega,B)/T(\omega,0)$ or differential spectra
$T(\omega,B+\Delta B)/T(\omega,B-\Delta B)$. The differential
technique is more precise and helps to correct for possible
magnetic field induced changes in the response of the bolometer.

\begin{figure}[t]
      \scalebox{0.43}{\includegraphics{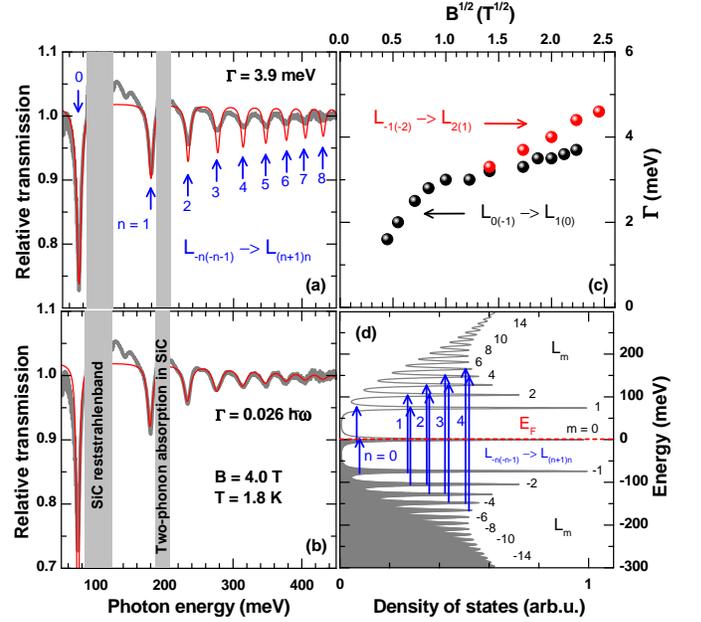}}
      \caption{\label{DOS} (color online) Part (a) and (b): Experimentally obtained transmission spectrum of multilayer
epitaxial graphene compared to theoretically expected curves.
(c) Broadening $\Gamma$ as derived from our data for L$_{0(-1)}\rightarrow$L$_{1(0)}$ and
L$_{-1(2)}\rightarrow$L$_{2(1)}$ transitions (d) DOS in graphene at $B=4.0$~T as calculated for the
energy-dependent broadening $\Gamma=0.026E$. The broadening of states around zero energy has been set  
constant to $0.026E_1$. Several observed absorption lines are schematically marked by vertical arrows.}
\end{figure}

A representative magneto-transmission spectrum at $B=4$~T
(normalized to the response at $B=0$) of our MEG sample is shown
in Fig.~\ref{DOS}(a),(b). This spectrum is composed of a series of
absorption lines which, accordingly to previous
studies~\cite{SadowskiPRL06,OrlitaPRL08II,JiangPRL07,DeaconPRB07,HenriksenPRL10},
%studies~\cite{SadowskiPRL06,OrlitaPRL08II,JiangPRL07},
can be easily identified as due to optically active transitions
between Landau levels $L_{\pm n}$ of massless Dirac fermions, with
the characteristic energies: $E_{\pm n}=\pm v_F\sqrt{2e\hbar
B|n|}$, $n=0, 1, 2\ldots$ These are the interband transitions:
L$_{-(n+1)}$$\rightarrow$L$_n$ and L$_{-n}$$\rightarrow$L$_{n+1}$,
which are reflected in the spectra of our quasi-neutral graphene
sheets. From the characteristic field dependence of these
transitions, $\hbar\omega=v_F \sqrt{2e\hbar
B}(\sqrt{n}+\sqrt{n+1})$, we find $v_F=(1.025\pm 0.005)\times
10^{6}$~m.s$^{-1}$, in agreement with previous estimations~\cite{SadowskiPRL06,OrlitaPRL08II}.
The L$_{-1(0)}$$\rightarrow$L$_{0(1)}$ transition disappeared from spectra (due to occupation effect at LL filling factor $\nu \approx 6$)
at fields of 100~mT which implies density $n\approx 10^{10}$~cm$^{-2}$ in the studied layers.

For a more quantitative analysis of the measured
magneto-transmission $T(\omega,B)$, we first note that the
absorption of light by our graphene sheets is relatively weak and
therefore:
\begin{equation}
T(\omega,B)=1-C\cdot\mathrm{Re}[\sigma_{xx}
(\omega,B)]/(\epsilon_0 c)
\label{transmission}
\end{equation}
where the proportionality factor
$C=N_{\mathrm{eff}}\cdot\frac{1}{2}(\kappa^{2}+3)/(\kappa^{2}+1)$
accounts for the effective number $N_{\mathrm{eff}}$ of graphene
sheets in our sample including the effect of possible partial
coverage of the SiC surface, and for multi-reflections of light
from the SiC substrate with the refractive index
$\kappa = 2.6$~\cite{SadowskiPRL06}. $\mathrm{Re}[\sigma_{xx}
(\omega,B)]$ is the real part of the dynamical longitudinal conductivity
of a single graphene sheet expressed in Kubo-Greenwood formalism~\cite{GusyninPRL07}:
\begin{equation}
\mathrm{Re}[\sigma_{xx}(\omega,B)]=\frac{e^{3}}{2\hbar}
\frac{B}{\omega} \sum_{r,s} M_{r,s}(f_s-f_r)\delta_\Gamma(E_r-E_s-\hbar\omega),
\label{conductivity}
\end{equation}
where $0 \leq f_r \leq 1$ stands for the occupation factor of the
$r^{\mathrm{th}}$ LL and $\delta_\Gamma(E)$ for a Lorentzian of width $\Gamma$.
$M_{r,s}=v_F^2\gamma\delta_{|r|,|s|\pm1}$ are velocity operator matrix elements
in which $\gamma=2$ if $r=0$ or $s=0$ and otherwise $\gamma=1$.

Equation \eqref{conductivity}
is a simplified version of a more general Kubo formula (see e.g. Refs.~\onlinecite{YangPRB10} and \onlinecite{YangPRB10II})
which includes initial and final spectral functions. Our simplified formula can be strictly derived
for energy-independent and purely imaginary self-energies of the initial and final LLs,
$i\Gamma_s$ and $i\Gamma_r$, and in the limit of relatively narrow
levels/transitions~\cite{GusyninPRL07}, i.e., $\Gamma_{r,s} \ll \hbar\omega$ and $\Gamma_{r,s}\ll |E_{r,s}|$. On the other hand,
since we deal with interband excitations only, the condition $\Gamma_{r,s}\ll |E_{r,s}-E_{r+1,s+1}|$ does not have to be fulfilled and adjacent LLs
thus may overlap. With these approximations, the broadening $\Gamma$ in Eq.~\ref{conductivity} corresponds to $\Gamma=\Gamma_r+\Gamma_s$,
and in a view of remarkable electron-hole symmetry of bands in graphene, also corresponds to $\Gamma\approx2\Gamma_{r,s}$.
At a given magnetic
field, the broadening $\Gamma$ can be viewed as a function of the resonance energy/frequency $\Gamma=\Gamma(\hbar\omega_{s \rightarrow r})$,
where $\hbar\omega_{s \rightarrow r}=|E_r|+|E_s|\approx2|E_{r,s}|$. In further, we also accept the possibility that $\Gamma=\Gamma(\hbar\omega)$ is a continues
function of the energy/frequency. As a matter of fact, the conductivity \eqref{conductivity} is practically identical if $\Gamma(\hbar\omega_{r \rightarrow s})$
is fixed for each transition or $\Gamma(\hbar\omega)$ is a slowly varying function over the interval $\hbar\omega_{r \rightarrow s}\pm\Gamma(\hbar\omega_{r \rightarrow s})$. 
This latter condition as well as $\Gamma_{r,s} \ll \hbar\omega$ and $\Gamma_{r,s}\ll |E_{r,s}|$ are a posteriori fulfilled in our experimental data.

Since we probe here
quasi-neutral graphene layers at low temperatures, we assume that the
Fermi energy is pinned to the $0^{\mathrm{th}}$ Landau level and
therefore $f_s=1$ and $f_r=0$ in the case of our interband
($r,s\neq0$), L$_s\rightarrow$L$_r$ transitions. As for a pair of
degenerate transitions, L$_{-1}\rightarrow$L$_0$ and
L$_0\rightarrow$L$_1$, which involve the 0$^{\mathrm{th}}$~LL, we
can, without loss of generality, set $f_1=0$ and $f_0=f_{-1}=1$
for any magnetic field, since the sum of intensities of
L$_{-1}\rightarrow$L$_0$ and L$_0\rightarrow$L$_1$ transitions is
independent of \textit{de facto} partial population of the
$L_{0}$ level ($|\nu|<2$)~\cite{OrlitaSST10}. The phenomenologically introduced
broadening parameter $\Gamma$ is the main focus of this work. In general,
this parameter depends on the magnetic field, on the transition
index, or, as discussed above on frequency $\omega$. Notably, the physical interpretation
of $\Gamma$ is different in the case of well-overlapped or well-resolved LLs.

In the limit of well separated Landau levels, $\Gamma= \Gamma_{r}+
\Gamma_{s}$ may correspond to a single-particle quantum level
broadening $\Gamma_{s}$ of the density of states (DOS)
\cite{GusyninPRL07}. This regime has been theoretically explored
in the case of long-range and short-range disorder, especially for scattering
centers insensitive to carrier concentration~\cite{ShonJPSJ98}.
For short-range scatterers, $\Gamma_{s}$ was found to be independent of the LL index
but to increase as $\sqrt{B}$. In the more complex case of
long-range scattering, $\Gamma_{s}$ may also vary as $\sqrt{B}$
but the broadening of the $0^{\mathrm{th}}$ LL is found
to be significantly (by a factor $\sqrt{2}$) larger than that of any
other levels. Charged impurities, if present, lead to a narrowing of LLs. The
width of the LL is then predicted to oscillate with the
filling factor and shows an overall $1/B$ field dependence,
$\Gamma \sim 1/ D(E_{F})$, instead of the $\sqrt{B}$ field
dependence~\cite{YangPRB10,YangPRB10II}.
As concluded by the authors of Ref.~\onlinecite{YangPRB10},
this excludes charge impurities as the dominant source of scattering
in MEG, in which $\sqrt{B}$-type broadening is characteristic
of inter-LL transitions~\cite{OrlitaPRL08II}.

In the limit of densely spaced and therefore strongly overlapping LLs, attained
in graphene at sufficiently high energies  -- see Fig.~\ref{DOS}(d) for illustration --  the
broadening parameter $\Gamma$ can be interpreted as the inverse of a transport relaxation time, $\Gamma=\hbar/\tau$
within a factor of 2. This interpretation can be justified by the classical Drude approach to describe the carrier motion
in a magnetic field~\cite{PalikRPP70,Kono}, in which the magnetic field effects are apparent when the
cyclotron frequency ($\omega_c=v^2_FeB/\epsilon$ for graphene) is comparable to the carrier
relaxation rate $\tau^{-1}$. The transport relaxation time is the only parameter which accounts
for broadening effects in the dynamical conductivity response of graphene at zero magnetic field~\cite{AndoJPSJ02}
and very likely in the magneto-dynamical conductivity response in the classical limit when the density
of states is weakly modulated by the magnetic field~\cite{PyatkovskiyPRB11}. Notably, our assumption $\Gamma=\hbar/\tau$ is favored
by the electron-hole symmetry in graphene and, therefore, in neutral sheets likely identical relaxation
rates for electrons and holes at energies $+\epsilon$ and $-\epsilon$, respectively. The transport relaxation time
becomes comparable with the single-particle relaxation time if short-range scattering dominates~\cite{HwangPRB08},
which is, as shown later on, consistent with our data and which thus facilitates our interpretation.

\begin{figure}[b]
\scalebox{0.32}{\includegraphics*{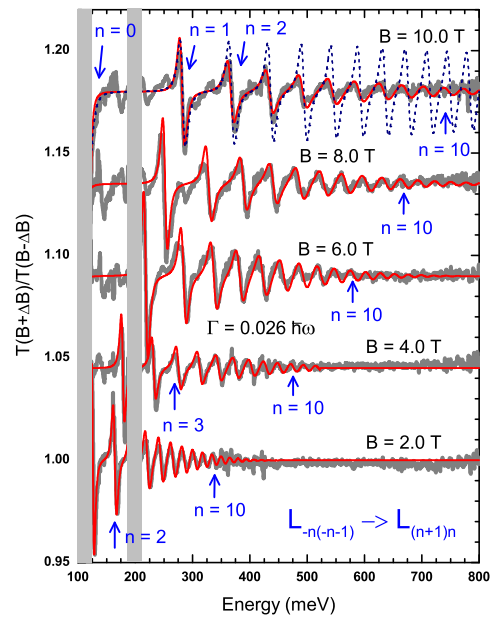}} \caption{\label{SPKT}
(color online) Differential transmission spectra $T(B+\Delta
B)/T(B-\Delta B)$. The lowest two spectra have been taken for
$\Delta B=0.05$~T, the other with $\Delta B=0.125$~T. For clarity, the
subsequent curves have been shifted by a factor of 0.045.
Experimental curves are accompanied by theoretical fits, assuming
the broadening parameter $\Gamma=0.026\hbar\omega$. Let us note that the condition
$B/\Delta B>2n$ must be fulfilled to prevent significant overlap
of adjacent transitions that involve $n^{\mathrm{th}}$ LL. The
dashed line shows the simulated curve for differential transmission at
$B=10$~T and $\Delta B=0.125$~T for a constant broadening
$\Gamma=8$~meV, which has been set to reproduce the
L$_{-1(-2)}\rightarrow$L$_{2(1)}$ line.}
\end{figure}

To compare with experiment, we re-consider the (relative)
magneto-transmission spectrum measured at $B=4$~T
in Fig.~\ref{DOS}. To model this spectrum, we assume that
$\sigma_{xx} (\omega, B=0) = e^{2}/4\hbar$. This is also the
numerical result from Eq.~\ref{conductivity} in the limit of
strongly overlapping levels. Then following Eqs.~\ref{transmission}
and \ref{conductivity} we fit $T(\omega, B)/T(\omega,0)$ with
two adjustable parameters/functions C and $\Gamma(\omega)$. In the
first and simplest approach, we set $\Gamma$ to be constant,
independent of the transition energy. As shown in Fig~\ref{DOS}(a),
we then fail to reproduce the high energy part of the spectrum, but
can well reproduce the two lowest, best-separated transitions. We
find $C=1.1$ and $\Gamma = 3.9$~meV by best fitting the shape of
the L$_{-1(0)}\rightarrow$L$_{0(1)}$ transition.

However, most of the spectrum consists of overlapping transitions
(in fact with the exception of the two lowest-energy transitions) that
can be very well reproduced by assuming $\Gamma=\alpha \cdot \hbar
\omega$. The best values are $\alpha = 0.026$ and $C=1.3$ from the
fit shown in Fig.~\ref{DOS}(b). This latter value implies a surprisingly low effective
number of layers, $N_{\mathrm{eff}}\approx 2$, in the present sample. We
related this to a partial coverage of the SiC surface by
graphene sheets and to the partial AB-stacking, e.g. into bilayers recently
identified as a minor component in MEG~\cite{SiegelPRB10,OrlitaPRB11}. Perhaps
another reason could be an increased dielectric screening inside MEG~\cite{ProfumoPRB10}, which can reduce the
total absorption significantly.
Remarkably, the $\Gamma=0.026\hbar\omega$ rule is universal, in the sense that
it applies to spectra measured at any magnetic field within the
range of the present measurements. This is illustrated in
Fig.~\ref{SPKT} with simulations of differential spectra measured
in the region of overlapping $n \geq 1$ transitions, at five
distinct magnetic fields. The calculated traces, which accurately fit
the data, are obtained from Eqs.~\ref{transmission} and
\ref{conductivity} with a common set of parameters ($\alpha =
0.026$, $C=1.3$). Here we note that despite a clear overlap between transitions, each of them is still well-defined in 
a sense that the above discussed assumption of narrow absorption lines and of slowly varying $\Gamma$ is indeed well 
satisfied, $\Gamma=0.026\hbar\omega \ll \hbar\omega$.

To sum up, the broadening parameter $\Gamma$ seems to be constant for transitions between
well-separated levels, but tends to increases linearly with energy when levels start to overlap significantly.
This behavior remains unchanged up to the highest resolved transitions, i.e., those which
originate from practically merging LLs. This is shown in Fig.~\ref{DOS}(d), where the DOS
has been plotted for the experimentally derived parameters. As discussed above, in the regime of strongly
overlapping levels, the broadening of electronic states is governed dominantly by the zero-field relaxation rate and
the observed linear in energy increase of $\Gamma$ indicates a $1/\tau\sim|\epsilon|$ dependence. Interestingly,
the same conclusion can be drawn using a simple semi-classical argument: the number of resolved transitions
in our spectra is roughly magnetic field independent, i.e., the onset of LL quantization remains always pinned
to a certain level ($n_0$$\sim$10, see Fig.~\ref{SPKT}). The energy of this onset
moves to higher energies as $|\epsilon|\sim \sqrt{B}$ and the cyclotron frequency corresponding to this onset
thus follows a linear in energy dependence $\omega_c=v_F^2 eB/|\epsilon|\sim|\epsilon|$. If we employ the classical condition for
this onset, $\omega_c \tau \sim 1$, we conclude $1/\tau\sim|\epsilon|$.

Having established the $\hbar/\tau = \alpha|\epsilon|$ dependence,
let us crudely assume that it is not additionally affected by
changes in the carrier concentration. The Fermi energy independent conductivity
of ``our graphene'' is then $\sigma=\frac{2}{\alpha}\cdot\frac{e^2}{h}$ ($\sigma\approx 80 e^2/h$ for $\alpha=0.026$)
and we can speculate on a
low-temperature carrier mobility, $\mu=e v_{F}^2\tau/E_{F}$, in
such a model system at various charge densities. As an
order of magnitude, we obtain $\mu\approx10^6$~cm$^{2}$/(V.s) for
$n=10^{10}$~cm$^{-2}$, which is consistent with our recent
measurements on quasi-neutral
graphene~\cite{OrlitaPRL08II}. The mobility falls
down to $\mu\approx10^4$~cm$^2$/(V.s) for the carrier density as
high as $n=10^{12}$~cm$^{-2}$, what reflects fairly well typical
mobilities measured for charged, single-layer epitaxial graphene
on C-face of SiC or interface layer of
MEG~\cite{WuAPL09,LinAPL10}. Our hypothetical graphene with $n= 10^{13}$~cm$^{-2}$
would have a mobility of $\mu\approx10^3$~cm$^2$/(V.s), in
striking agreement with typical values measured for high
concentration graphene on Si-face of
SiC~\cite{EmtsevNatureMater09}.

The $1/\tau(\epsilon)$ dependence reported here is expected for
graphene and points towards conventional scattering mechanisms in
quasi-neutral epitaxial graphene. Apart from charge impurities,
excluded already earlier in MEG due to $\sqrt{B}$-broadening of inter-LL
transitions~\cite{YangPRB10}, we can also for certain exclude the
dominant role of resonant scatterers, which produce the mid-gap states and
therefore modify the predicted $1/\tau \sim |\epsilon|$ relation.
Those resonant scatters are by definition extrinsic, and maybe
important for environment exposed graphene, but are naturally expected to be
absent in our protected layers.

To complete our discussion of scattering mechanisms in MEG, we focus on the
two lowest energy transitions of Fig.~\ref{DOS}(a),(b). These are the
two best separated transitions in our spectra, apparently related to nearly
non-overlapping LLs ($|n|=0,1$ and 2) and therefore, their behavior
can be tentatively analyzed in reference to theory in Ref.~\onlinecite{ShonJPSJ98}.
Figure~\ref{DOS}(c) shows that the
width of both transitions scales roughly as $\sqrt{B}$. Moreover,
the width of the $n=0$ transitions is roughly the same if not even
smaller than the width of the $n=1$ transition. According to
the theoretical analysis of the dynamical magneto-conductivity in the
regime of well-separated Landau levels, this observation points
towards short- rather than long-range scattering mechanism
(both insensitive to carrier energy). Another possibility is the
electron-electron scattering which also provides a scattering rate that increases
linearly with energy~\cite{PoliniPRB08}.

Concluding, we have investigated the dynamical
magneto-conductivity of quasi-neutral epitaxial graphene
layers and studied the carrier scattering in these layers. We find
that the relaxation rate is linear with the carrier energy. This
dependence has been initially expected for graphene, nevertheless, not clearly observed yet.
Its interesting implication is that the number of resolved Landau levels is fixed,
independent of the magnetic field. The dominant scattering
mechanism in the extrinsically unperturbed layers of epitaxial graphene
is conventional, very likely due to short-range potentials or due to electron-electron
interactions.

\begin{acknowledgments}
We acknowledge funding received from EuroMagNETII under the EU
contract 228043, from the Keck Foundation, the NSF-MRSEC grant
0820382 from the Partner University Fund at the Embassy of France.
This work has been further supported by the projects RTRA DISPOGRAPH,
PICS-4340 \& 395/N-PICS-FR/2009/0, MSM0021620834 and GACR P204/10/1020, as
well as by grants 670/N-ESF-EPI/2010/0, 671/N-ESF-EPI/2010/0 and
GRA/10/E006 (EPIGRAT).
\end{acknowledgments}

%\bibliography{Scattering-time}

\end{document}